\documentclass[12pt]{report}
\usepackage{latexsym}
\usepackage{amsmath,amsfonts,graphicx}
\usepackage{amsbsy}
\usepackage{hyperref}
\usepackage{mathrsfs}
\usepackage{color}
\usepackage{psfrag}
\usepackage{enumerate}
\usepackage{amsmath,amssymb,calc,amsfonts}
\usepackage{latexsym}
\usepackage[utf8]{inputenc}
\DeclareFontFamily{U}{rsfs}{}         
\DeclareFontShape{U}{rsfs}{m}{n}{<5> rsfs5 <6><7> rsfs7          %
  <8><9><10><10.95><12><14.4><17.28><20.74><24.88> rsfs10}{}     %
\DeclareMathAlphabet{\mathfs}{U}{rsfs}{m}{n}                     %
    
\newcommand{\dbar}{d\hspace*{-0.08em}\bar{}\hspace*{0.1em}}                           %
\definecolor{indiagreen}{rgb}{0.07, 0.53, 0.03}
\def\beq{\begin{eqnarray}}
\def\eeq{\end{eqnarray}}

\def\g{\gamma}

\def\={\stackrel{\Delta}{=}}

\def\grad{\nabla}

\def\EH{Einstein-Hilbert }

\def\dbar{{\mathchar'26\mkern-12mu d}}

\begin{document}

\title{Black Hole Thermodynamics: General Relativity \& Beyond \vskip 0.5 cm  \large{Sudipta Sarkar \footnote{sudiptas@iitgn.ac.in}
\\ Indian Institute of Technology, Gandhinagar, 382355, Gujarat, India.}}

  \maketitle
  
    \begin{abstract}
 Black holes have often provided profound insights into the nature of gravity and the structure of space-time. The study of the mathematical properties of black objects is a major research theme of contemporary theoretical physics. This review presents a comprehensive survey of the various versions of the first law and second law of black hole mechanics in general relativity and beyond. The emphasis is to understand how these laws can constrain the physics beyond general relativity.
   \end{abstract}
   
  \textit{ ``The black holes of nature are the most perfect macroscopic objects there are in the universe: the only elements in their construction are our concepts of space and time." } \vskip 0.5 cm
   .... {\bf Subrahmanyan Chandrasekhar}

\section{Introduction}

It is appropriate to start a review of black hole thermodynamics with the above quotation by S. Chandrasekhar. The quote brings out the fundamental characteristics of a black hole: \textit{The Universality}. The properties of a black hole are (almost) independent of the details of the collapsing matter, and this universality is ultimately related to the fact that black holes could be the thermodynamic limit of underlying quantum gravitational degrees of freedom. Therefore, the classical and semi-classical properties of black holes are expected to provide important clues about the nature of quantum gravity. A significant obstacle in constructing a theory of quantum gravity is the absence of any experimental or observational result. The only ``test" we can imagine is the theoretical and mathematical consistency of the approach. The understanding of the fundamental laws of black hole mechanics could be a necessary (if not sufficient) constraint on the theory of quantum gravity. \\

The modern understanding of the properties of black hole starts with the resolution of the ``Schwarzschild Singularity" using Kruskal-Szekeres coordinates \cite{Kruskal:1959vx, Szekeres:1960gm}. These coordinates cover that entire spacetime manifold of the maximally extended vacuum spherically symmetric solution of the Einstein's field equation and are well-behaved everywhere outside the physical singularity at the origin, in particular at the position $r = 2M$. The next important step is the discovery of the rotating asymptotically flat vacuum black hole solution by Roy Kerr \cite{Kerr:1963ud}. The solution exhibited various interesting and generic properties of a stationary black hole in general relativity. The existence of Ergosphere and Superradiance show how to extract energy and angular momentum from the black hole. The study of these phenomena lead to a significant result; the area of the black hole can never be decreased using these processes. For example, using the Penrose process, it is possible to extract energy from the black hole, and as a result, the mass of the black hole decreases. At the same time, the process slowed down the rotation, and the net effect only increases the area. \\

Then comes the famous work by Hawking \cite{Hawking:1971vc} which analyzes the general properties of a black hole, independent of the symmetry of a particular solution. This work contains several important theorems: the topology theorem, the strong rigidity theorem and most importantly, the area theorem. Area theorem is a remarkable result which asserts that the area of the event horizon can not decrease as long as the matter obeys a specific energy condition. This is a highly nontrivial statement related to the dynamics of black holes in general relativity. Consider the collision of two black holes which generated a burst of gravitational waves extracting energy from the black holes to infinity. The area theorem constrains the efficiency of this process and limits the amount of radiated energy so that the area of the final black hole is always greater than the sum of the individual black hole areas before the collision \cite{Hawking:1971tu}. In this sense, the area theorem is a statement of the limitation of converting the black hole mass into energy; akin to the second law of thermodynamics.\\

Immediately after this result, Hawking, Bardeen, and Carter wrote down the four laws of black hole mechanics \cite{Bardeen:1973gs} and as it is well known, these laws have an intriguing resemblance with the laws of thermodynamics. Interestingly, the paper treated this resemblance as only a formal analogy. The real step towards the black hole thermodynamics was taken by Bekenstein, \cite{Bekenstein:1972tm, Bekenstein:1973ur} who proposes that we should take the area theorem seriously and the area of the black hole is indeed related to the thermodynamic entropy of the event horizon. The basis of this claim was somewhat heuristic; Gedanken experiments estimating the loss of information due to the presence of the horizon. The arguments show that the entropy is proportional to the area of the event horizon and therefore the area theorem is a consequence of the second law of thermodynamics. These results mark the real beginning of the black hole thermodynamics, and the analogy becomes a robust correspondence with the discovery of the Hawking radiation \cite{Hawking:1974sw} fixing the proportionality constant between area and entropy. The final expression of the Hawking temperature and Bekenstein entropy of the horizon of a $4-$dimensional Schwarzschild black hole of mass $M$ and area $A$ becomes;

\begin{equation}\label{temanden}
T_H =\frac{ \hbar\, c^3 }{8 \pi G k_B\, M}; \hskip 0.5 cm \,\,\, S = \frac{c^3 k_B\, A}{4 G \hbar },
\end{equation}

It is evident from the appearance of the Planck constant, and the Newton's constant, the understanding of these expressions would require some form of quantum gravity. It may be possible to proclaim that these are the leading order result of the theory of quantum gravity \footnote{As a side remark, let me point out that the Chandrasekar mass formula also contains all the fundamental constants. But we do not associate that result with the quantum gravity.}. \\

This review aims to understand the issue of the general applicability of the laws of black hole thermodynamics. In particular, we will try to answer the following questions;
\begin{itemize}
\item How far the laws of Black Hole mechanics can be generalized beyond General Relativity? 
\item Can we constrain possible extensions of general relativity using the black hole (BH) mechanics?
\item What exactly we have learned so far about Quantum Gravity from BH Mechanics?
\end{itemize}

The last question is indeed the toughest and probably remain unanswered in this review except for some rudimentary remarks at the end. But, given the recent developments of black hole physics, it is possible to provide reasonable answers to the first two questions. \\

The discussion in the review will be mostly classical, and we will assume the applicability of the classical energy conditions, in particular, the null energy condition. The primary focus is a comprehensive discussion of the physical process law and the second law.  We will not consider the issues related to the semiclassical gravity; in particular, Hawking radiation and trans Planckian problem. Another vital omission will be the information loss paradox. We will also restrict ourselves to the mechanics and thermodynamics of the event horizon only.

\section{The various versions of the first law}

The first law of black hole mechanics has several avatars, and we need to distinguish the different formulations of the first law. In ordinary thermodynamics, the first law is the statement of the conservation of energy. The total energy can not be destroyed or created, but can always be converted into another form of energy. The statement is mathematically described by the difference equation $\Delta U  = Q - W$. The change of the internal energy $U$ of the system is equal to the difference of the heat supplied $Q$, and the work done $W$ by the system. The conservation of energy is built-in into the dynamics of general relativity. So, what we mean by the first law is the Clausius theorem which involves the notion of the entropy. Consider a system under quasi-static change which is subjected to an infinitesimal amount of heat $\dbar Q$ from the surrounding. The heat change is an inexact differential, and therefore the total heat $Q$ is not a state function. It is then assured that there exists a state function called the ``entropy" $S$ such that the temperature of the system acts as an integrating factor relating the change in entropy to the heat supplied as $ T dS = \dbar Q$. The Clausius theorem ensures the existence of the state function entropy associated with a thermodynamic equilibrium state of the system. Note that all changes are considered to be quasi-stationary, always infinitesimally close from an equilibrium state.\\

In the case of a black hole, we need to be careful before applying these concepts. To begin, the obvious choice of an equilibrium state is a stationary black hole. So, let us first define the notion of the stationary black hole in general relativity. \\

To define the event horizon of a black hole, we require information about the asymptotic structure. Suppose we consider an asymptotically flat space-time such that the asymptotic structure is the same as that of the flat space-time. Then, the event horizon is defined as the complement of the past of the future null infinity. This is a global definition and to find the location of the event horizon, we require the knowledge about the entire space-time. This is not a very convenient concept. For example, if one is looking for the signature of the formation of the event horizon in the computer codes of numerical relativity, she has to wait for infinite time! As a result, alternative notions like apparent horizon and quasi-local horizons may suit much better for such an analysis. Nevertheless, the event horizon can be very useful because it is a null surface \footnote{A proof that the event horizon in an asymptotically flat spacetime is a null surface is given in \cite{Hawking:1971vc} and also discussed in detail in \cite{Townsend:1997ku} }, and the causal boundary between two regions of space-time called inside and outside of the black hole. As a result, at least intuitively it makes sense to assign an entropy to the null event horizon. \\

The definition of the event horizon does not need any symmetries of the underlying space-time. Now, consider the particular case when the space-time is stationary and contains a time like Killing vector. Such a time like Killing vector provides a related concept called the Killing horizon. A Killing horizon is a surface where the time like Killing field becomes null. An example of such a surface would be the Rindler horizon in the flat space-time. It is easy to check that the boost Killing field indeed becomes null at the location of the Rindler horizon. This example shows that the Killing horizon may be entirely unrelated to the event horizon. The Rindler accelerated horizon is a Killing horizon but not an event horizon.\\

Next, consider an event horizon in a stationary space-time. Then, it is the Strong Rigidity theorem \cite{Hawking:1971vc} which asserts that the event horizon in a stationary space-time is also a Killing horizon. The strong rigidity theorem is a powerful result, and the proof requires Einstein field equation and some technical assumptions like the analyticity of the space-time. Generalizing the proof beyond $3+1$ dimensions needs more sophisticated mathematical machinery \cite{Hollands:2006rj, Moncrief:2008mr}. The strong rigidity theorem is only proven for general relativity. Therefore for black holes in various modified gravity theories, we have to consider this as an assumption. \\

The derivation of the equilibrium state version of the first law starts with a stationary event horizon which is also a Killing horizon. For simplicity, the $D$ dimensional spacetime is assumed to be asymptotically flat. We will also assume that the physical space-time can be extended to add a bifurcation surface in the past, where the time like Killing field vanishes. The existence of a bifurcation surface ensures that the surface gravity is constant along all the directions on the horizon \cite{Kay:1988mu}. We will consider that the bifurcation surface is regular, i.e., all the fields have a smooth limit from the outside to the bifurcation surface. This is a nontrivial assumption, and there are theories, e.g., Einstein-Aether Theory \cite{Foster:2005fr} in which such an assumption does not hold.  Given all these, we now write down the expression of the ADM mass, or in this case the Komar mass as,

\begin{equation}\label{def_Komar}
M = -\frac{1}{8 \pi} \int_{S} \nabla^a \xi^b \, dS_{ab}.
\end{equation}
The integration is at the asymptotic spatial infinity, and the Killing field is normalized as $\xi_a \xi^a = -1$ asymptotically. For the black hole spacetime, let us consider a space like hypersurface $\Sigma$ which extends from infinity to the horizon. The surface has two boundaries, one at infinity and other at the horizon. Using Stokes theorem and Einstein's field equations $G_{ab} = 8 \pi T_{ab}$, we can then express the Komar mass as,

\begin{equation}
M = -2 \int_{\Sigma} \left(T^{a}_{b} \xi^b - \frac{1}{D - 2} T \xi^a \right) d\Sigma_a + \frac{1}{8 \pi} \int_{H} \nabla^a \xi^b \, dS_{ab},
\end{equation}

where $T$ denotes the trace of the energy-momentum tensor. Let us further assume that we are only considering a vacuum solution; and therefore $T_{ab} = 0$.  Also, there is no angular momentum, and the space-time is static. Then the first integral vanishes. The last integral is at the inner boundary of the surface $\Sigma$ where it meets the horizon. For the static spacetime, we can evaluate the second integral and write the final expression as \cite{Bardeen:1973gs},

\begin{equation}
M = \left(\frac{D - 2}{D - 3}\right) T_H S,
\end{equation}

where $T_H$ is the Hawking Temperature and $S$ is the Bekenstein entropy of the black hole. The expressions of $T_H$ and $S$ contain the Planck constant $\hbar$, but the product is independent of $\hbar$. This equation is a particular case of what is known as the Smarr formula \cite{Smarr:1972kt}. Although the derivation of this equation is straightforward, the interpretation is a bit subtle. The equation relates an asymptotic quantity, the ADM/Komar mass with the quantities corresponding to the horizon. If we approve the use of thermodynamic concepts, the Smarr formula may be regarded as the equation of state at thermodynamic equilibrium relating energy $M$, temperature $T_H$ and entropy $S$. Also, the derivation does not work for $D = 3$ indicating the absence of asymptotically flat vacuum black hole solutions in lower dimensions. Note that, there is no physical process by which we can change the ADM or Komar mass. This formula is only valid for a strictly static and vacuum space-time. Therefore, instead of a physical change, let us now consider a virtual change of the quantities: two Schwarzschild black hole solutions in $D$ dimensions with masses $M$ and $ M + \Delta M$ in the space of solutions of general relativity. Therefore the variation $\Delta M$ represents a virtual change, again only in the space of static, vacuum and asymptotically flat black hole solutions of general relativity. Then, the variation of the Smarr formula gives,

\begin{equation}
\Delta M = \left(\frac{D - 2}{D - 3}\right) \left(T_H \Delta S + S \Delta T_H \right).
\end{equation}
Let us evaluate the $r.h.s$ of the above equation for a Schwarzschild black hole in $D$ dimensions. Set $\left(G = \hbar = k_B = c = 1\right)$ and then the metric is  ,
\begin{equation}
ds^2 = -\left( 1- \frac{C}{r^{D-3}}\right) dt^2 + \frac{dr^2}{\left( 1- \frac{C}{r^{D-3}}\right)}+r^2d\Omega^2.
\end{equation}
The constant $C$ is a function of the ADM mass $M$ of the space time and if $D=4$, we have $C=2 M$. The horizon is located at $r_h = C^{\frac{1}{D-3}}$ and the surface gravity is $\kappa = ((D-3)/2)C^{-\frac{1}{D-3}}$. The expression of the Hawking Temperature and the Bekenstein entropy are then given by,

\begin{equation}\label{temanden}
T_H =  \frac{\kappa}{2 \pi} = \frac{D-3}{4 \pi} C^{-\frac{1}{D-3}}; \hskip 0.5 cm \,\,\, S = \frac{A_{D-2} C^{\frac{D-2}{D-3}}}{4 }.
\end{equation}
Using these expressions, it is easy to verify that,
\begin{equation}
\Delta M = \left(\frac{D - 2}{D - 3}\right) \left(T_H \Delta S + S \Delta T_H \right) = T_H \Delta S,
\end{equation}
This is the simplest derivation of the \textit{equilibrium state version} of the first law of black hole mechanics. 

This derivation can be generalized in several ways. If we include matter, e.g., an electrovacuum solution,  there will be additional work terms. But the most interesting generalization is for theories with higher curvature terms in the action. The area law fails generically for higher curvature gravity \cite{Visser:1993qa, Jacobson:1993xs, Visser:1993nu, Wald:1993nt, Iyer:1994ys} and the entropy is proportional to a different local geometric quantity evaluated on the horizon. In fact, the black hole entropy in any diffeomorphism invariant theory of gravity turns out to be the Noether charge of the Killing isometry which generates the horizon \cite{Wald:1993nt, Iyer:1994ys}. Before discussing the derivation of this ``Wald entropy", we will first try to understand intuitively why and how the area law fails beyond general relativity, using a generalized version of the original argument by Bekenstein  \cite{Bekenstein:1972tm, Bekenstein:1973ur}.

There are several motivations of considering a higher curvature theory. As a typical example, consider the perturbative quantization of gravity which leads to nonrenormalizable quantum theory and is confronted by uncontrollable infinities. If we treat such a nonrenormalizable theory as a low-energy effective field theory, adding new counter-terms and couplings at each new loop order, then the effective Lagrangian of gravity can be expressed as \beq
{\cal L} = \frac{1}{16 \pi G} \left( R + \alpha \, {\cal O}(R^2) + \beta \, {\cal O}(R^3) + \cdots\right),
\eeq
where $\alpha, \beta, \cdots$ are the new parameters of the theory with appropriate dimensions of length.  At the level of the effective theory, all terms consistent with diffeomorphism invariance can appear, but from a phenomenological point of view, only a subset of terms which leads to a well behaved classical theory is more desirable. In this case, the motivation of having these higher curvature terms comes from the idea that the Einstein-Hilbert action is only the first term in the expansion for the low energy effective action and higher order terms arise from the quantum corrections to the Einstein-Hilbert action functional \cite{Deser:1974cz}, which will, of course, depend on the nature of the microscopic theory. In particular, such higher curvature terms also arise in the effective low energy actions of some string theories \cite{ Zwiebach:1985uq, Boulware:1985wk}.

The detailed structure of these terms will depend on the specifics of the underlying quantum gravity theory. If we turn on these higher curvature corrections, the field equation will get modified, and the area theorem may not hold anymore. But, for specific higher curvature terms, we can still obtain exact black hole solutions as in case of GR. Now, consider the simplest case of spherical symmetry and assume that a set of identical particles with the same mass $m$ is collapsing in $D$ dimensions to form a black hole of mass $M$. If each of these particles contains one bit of information (in whatever form, may be information about their internal states, etc.), then the total loss of information due to the formation of the black hole will be $ \sim  M/m$. Classically, this can be as high as possible, but quantum mechanically there is a bound on the mass of each constituent particle because we want the Compton wavelength of these particles to be less than the radius of the hole $r_h$. Then, the maximum loss of information will be $\sim  M r_h$, and this is a measure of the entropy of the hole. Note that, we have not used any information about the field equation yet. So, this is completely an off-shell result. The field equation will provide a relationship between the mass $M$ and the horizon radius.\\
Let us now treat the specific case of general relativity. If we solve the vacuum Einstein's equations for spherical symmetry, we obtain the usual Schwarzschild solution with $M \sim r_{h}^{D-3}$, and this lead to black hole entropy proportional to $r_{h}^{D-2}$, the area of the horizon. \\
Next comes the modified gravity, with higher curvature terms and we will have new dimensionful constants in our disposal. Therefore, there could be a complicated relationship between mass and horizon radius. For example, if we restrict ourselves up to only curvature square correction terms with a coupling constant $\alpha$, we could have a relationship like $ M \sim r_{h}^{D-3} + \alpha \, r_{h}^{D-5}$, and the second term results in a sub-leading correction to black hole entropy. This simple illustration shows how the presence of new dimensionful constants in modified gravity theories leads to a possible modification of the black hole entropy.\\

The simplest way to derive the first law for any higher curvature theory would be to start with a suitable modification of the definition of the Komar mass in Eq.~(\ref{def_Komar}). For example, if we are working with $m$-th Lovelock class of action functionals with Lagrangian ${\cal L}^{(m)}$, the appropriate definition of the Komar mass will be \cite{Kastor:2008xb},

\begin{equation}\label{def_Komar_Lovelock}
M = -\frac{1}{8 \pi} \int_{S} P_{abcd} \nabla^c \xi^d \, dS_{ab},
\end{equation}
where the tensor $P_{abcd}$ has the symmetries of the Riemann Curvature tensor and is defined as,

\begin{equation}
 P^{abcd} = \frac{\partial {\cal L}^{(m)}}{\partial R_{abcd}}.
\end{equation}

For Lovelock gravity, the tensor also has the property $\nabla_i P^{abcd} = 0; \,\, i = a, b, c, d$. Using this expression and also the properties of the Killing vector, it is possible to derive a Smarr formula \cite{Liberati:2015xcp} exactly as in case of GR but the entropy as;

\begin{equation}
 S_w = - 2\pi \int_{{\cal B}} P^{abcd} \epsilon_{ab} \epsilon_{cd} \sqrt{h} \,d^{D-2} x, \label{waldf}
\end{equation} 
where $\epsilon_{ab} $ is the bi-normal to the bifurcation surface ${\cal B}$. As in the case of general relativity, the entropy obeys a Clausius theorem $ T \Delta S = \Delta M$ for infinitesimal variation in the space of static vacuum solutions.  \\

This simple derivation can be made more rigorous by using the Noether charge formalism of Wald and collaborators \cite{Visser:1993qa, Jacobson:1993xs, Visser:1993nu, Wald:1993nt, Iyer:1994ys}. The crucial input to the derivation is the diffeomorphism invariance in the presence of an inner boundary. The bulk part of the Hamiltonian vanishes on-shell, and the two boundary terms (one at the horizon and other at the outer boundary) are related to each other. Then for variations in the space of stationary solutions, we get the first law as the Clausius theorem. \\
The construction of Wald entropy formula crucially depends on the existence of a bifurcation surface. However, as pointed out by \cite{Jacobson:1993vj}, the Wald entropy remains unaffected even when it is evaluated on an arbitrary cross-section of a stationary event horizon provided the bifurcation surface is regular. The Noether charge construction also has several ambiguities, but,  the ambiguities in the Noether charge construction doesn't affect the Wald entropy in case of stationary black holes \cite{Iyer:1994ys, Jacobson:1993vj}. However, if the horizon is involved in a dynamical process, i.e., for nonstationary black holes, the Wald entropy formula no longer holds and turns out to be ambiguous up to the addition of terms proportional to the expansion and shear of the dynamical event horizon. \\
\begin{figure}[h!]
\begin{center}
\includegraphics[scale=0.6]{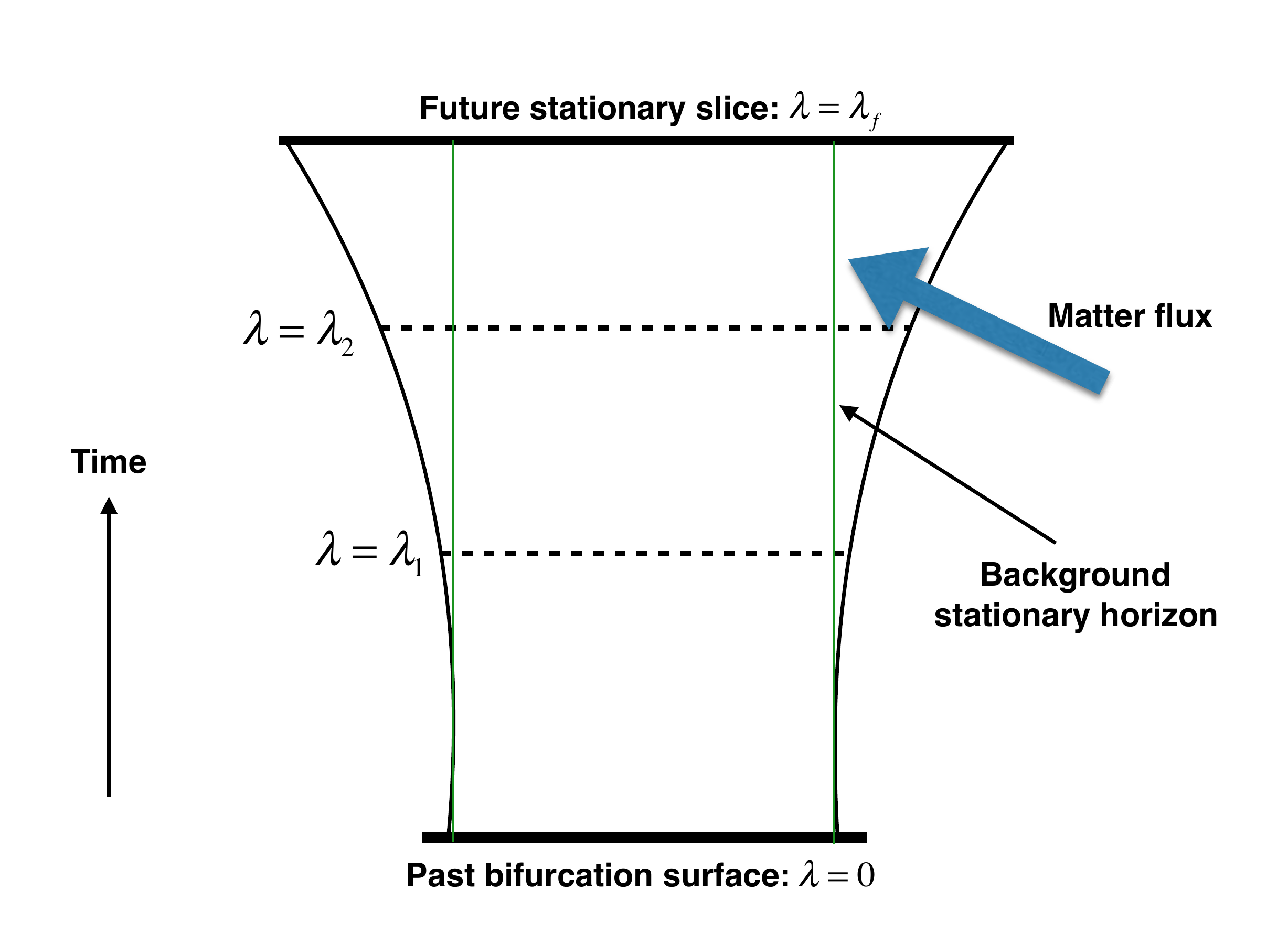}
\caption{A pictorial depiction of the geometry considered in the Physical process version of the first law. The green line depicts the evolution of the unperturbed stationary event horizon while the black curve denotes the evolution of the perturbed dynamical event horizon. The change in area is calculated between the two slices $\lambda=0$ (the bifurcation surface) and $\lambda=\lambda_f$ (a stationary final slice) along the black curve. }
\label{fig.PP}
\end{center}
\end{figure}
\\
Having discussed the equilibrium state version of the first law, let us now focus on another version of the first law for black holes: \textit{The physical process law}. This version of the first law involves the direct computation of the horizon area change when a flux of matter perturbs the horizon \cite{Hawking:1972hy, Wald:1995yp, Jacobson:2003wv} (henceforth referred to as PPFL). Unlike the equilibrium state version, PPFL is local and does not require the information about the asymptotic structure of the space-time and is therefore expected to hold for a wide class of horizons (see Figure: Eq.~(\ref{fig.PP})). Consequently, after some initial debate regarding the applicability of PPFL in the context of Rindler space-time  \cite{Jacobson:2003wv}, it was later demonstrated, following  \cite{Amsel:2007mh, Bhattacharjee:2014eea}, that the physical process version of first law indeed holds for Rindler horizon in flat space time, or for that matter, any bifurcate Killing horizon. 

Consider a situation, in which a black hole is perturbed by matter influx with stress-energy tensor $T_{ab}$ and it finally settles down to a new stationary state in the future. Then the PPFL which determines the change of the horizon area $A_{\rm H}$ is given by,
\begin{align}\label{BH_Dyn_01}
\frac{\kappa}{2\pi}\delta \left(\frac{A_{\rm H}}{4}\right)=\int_{{\cal H}}  T_{ab}\,\xi^{a}\, d\Sigma^b~.
\end{align}
Here $d\Sigma^b = k^{b} \,dA\,d\lambda$ is the surface area element and $k^a = \left( \partial / \partial \lambda \right)^a$ stands for the null generator of the horizon. The integration is over the dynamical event horizon and the affine parameter $\lambda$ varies from the bifurcation surface (set at $\lambda = 0$) to the future stationary cross section at $\lambda = \lambda_f$. Also, the background event horizon is a Killing horizon with the Killing field $\xi^a$ being null on the background horizon. On the background horizon surface, it is related with the affinely parametrized horizon generator ($k^a$) as $\xi^a = \lambda \kappa k^a$, where $\kappa$ is the surface gravity of the background Killing horizon. It is important to note that the derivation of the above result crucially hinges on the fact that the terms quadratic in expansion and shear of the null generator $k^{a}$ can be neglected since the process has been assumed to be sufficiently close to stationarity. This approximation ensures that there will be no caustic formation in the range of integration.\\

For PPFL, the variation $\delta A_{\rm H}$ represents the physical change in the area of the black hole due to the accretion of matter. As a result, here we are considering a genuinely dynamical situation. The physical process first law, therefore, relates the total change of entropy due to the matter flux from the bifurcation surface to a final state. If we assume that the black hole horizon is stable under perturbation, then the future state can always be taken to be stationary with vanishing expansion and shear, and the initial cross-section can be set at the bifurcation surface ($\lambda=0$). The choice of these initial and final states are necessary for this derivation of the physical process first law, to make some boundary terms vanish. The derivation can be generalized to obtain the expression of the entropy change between two arbitrary nonequilibrium cross sections of the dynamical event horizon. The additional boundary terms appearing in Eq.~(\ref{BH_Dyn_01}) are then related to the energy of the horizon membrane arising in the context of the black hole membrane paradigm \cite{Chakraborty:2017kob}. 

To elaborate on the derivation of the PPFL, let us start by describing the horizon geometry and set up the notations and conventions. We will follow the derivation as presented in \cite{Chakraborty:2017kob}.

\section{General Structure of PPFL}\label{GSPPFL}

\begin{figure}
\centering
\includegraphics[scale=0.6]{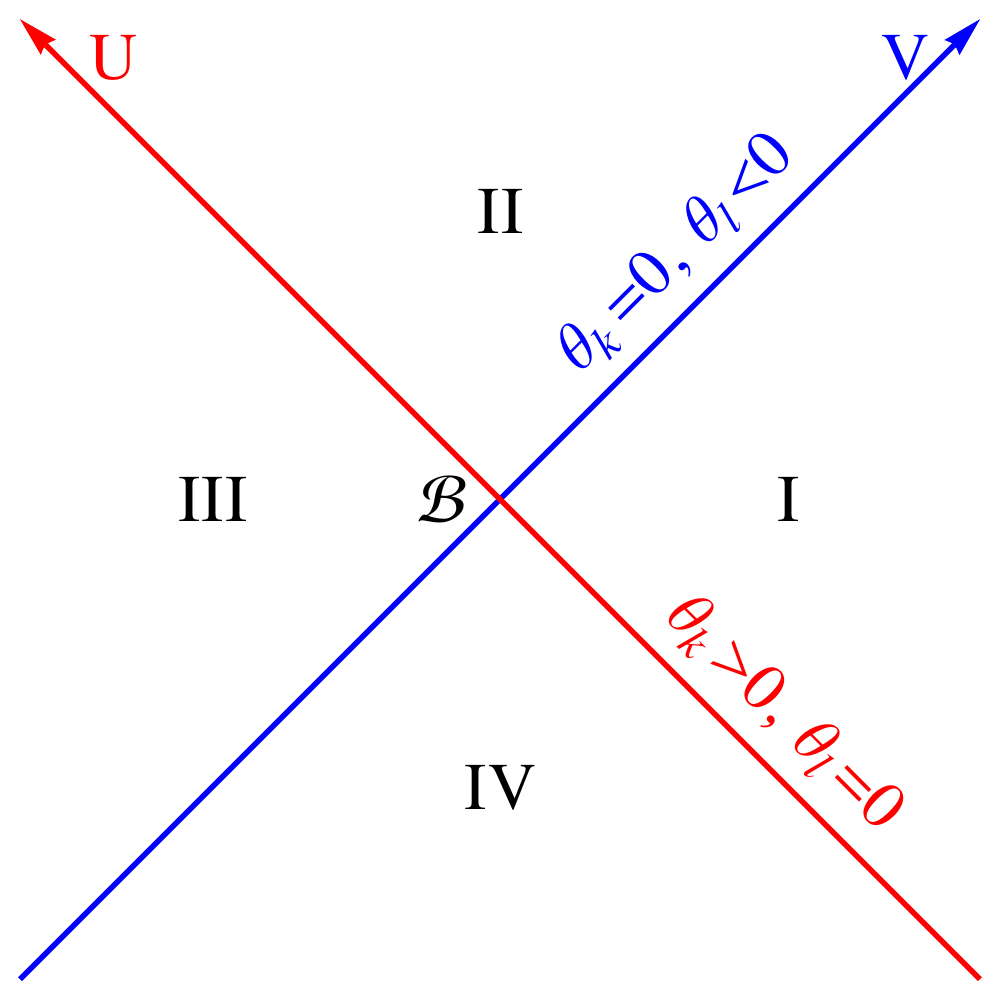}
\caption{The point $\mathcal{B}~(U=0,V=0)$, a $(D-2)$-dimensional cross-section of the Horizon represents the Bifurcation surface, where $\theta_k,\theta_l=0$.}\label{Kruskal}
\end{figure}

The event horizon $H$ of a stationary black hole in $D$ spacetime dimensions is a null hypersurface generated by a null vector field $k^a=(\partial / \partial\lambda)^a$ with $\lambda$ being an affine parameter. The cross section ($\mathcal{H}$) of the event horizon, which is a co-dimension two, spacelike surface, can be taken as $\lambda = \textrm{constant}$ slice. Being a co-dimension two surface, $\mathcal{H}$ posses two normal direction. One of them is the null normal $k^{a}$ and the other corresponds to an auxiliary null vector $l^{a}$ defined on $\mathcal{H}$ such that $k_a l^a =-1$. Then, the induced metric on the horizon cross section takes the form, $h_{ab} = g_{ab}+k_a l_b + k_b l_a$. Taking $x^A$ to be the coordinates on $\mathcal{H}$, $(\lambda,x^A)$ spans the horizon. 
We define the expansion and shear of the horizon to be the trace and traceless symmetric part of the extrinsic curvature and denoted as $(\theta_k,\sigma^k_{ab})$ and $(\theta_l,\sigma^l_{ab})$ with respect to $k^a$ and $l^a$ respectively. Taking $h$ to be the determinant of the induced metric $h_{ab}$, the expansion $\theta_k$ of the horizon can be written as,
\begin{equation}
\theta_k = \frac{1}{\sqrt{h}} \frac{d}{d\lambda}\sqrt{h}.
\end{equation}
Then, the evolution of $\theta_k$ along the horizon with respect to the affine parameter $\lambda$ is governed by the Raychaudhuri equation,
\begin{equation}
\frac{d\theta_k}{d\lambda} = -\frac{1}{D-2}\theta_k^2 -\sigma^k_{ab}\sigma^{ab}_k - R_{ab}k^a k^b\label{rc}.
\end{equation} 
As mentioned before, an important notion that will play a significant role throughout our discussion is the bifurcation surface. A bifurcation surface is a $(D-2)$ dimensional spacelike surface $\mathcal{B}$, on which the Killing field $\xi^a$ identically vanishes. Also $\mathcal{B}$ is the surface on which the past and future horizons intersect. For our purpose, it is convenient to choose $\mathcal{B}$ to be at $\lambda=0$. This choice can always be made due to the freedom to choose the parametrization of the horizon. The bifurcation surface is not a part of black hole spacetime formed by the gravitational collapse of an object. However, if the geodesics that generate the horizon are complete to the past, one can always have a bifurcation surface at some earlier $\lambda$. This can be realized by the maximal extension of the black hole space-time. For instance, no notion of bifurcation surface exists in the Schwarzschild space-time. Nevertheless, in its maximal extension, i.e., in the Kruskal space-time, the $2$-sphere at $U=0, V=0$ represents a bifurcation surface, as indicated in Eq.~(\ref{Kruskal}). A simple calculation leads to the following expressions of the expansion coefficients along $k$ and $l$,
\begin{equation}
\theta_k \propto U;\qquad \theta_l \propto V.
\end{equation}
Hence, at the future horizon $U = 0$, the expansion $\theta_k = 0$ and at the past horizon where $V = 0$, we have $\theta_l = 0$. At the bifurcation surface $(U=0,V=0)$ both $\theta_k$ and $\theta_l$ vanishes. Also, the shears can be shown to be vanishing on $\mathcal{B}$. 

$\theta_k$ and $\sigma_k$ are of first order in perturbation, i.e., $\mathcal{O}(\epsilon)$, while $\theta_l$ and $\sigma_l$ are of zeroth order, with $\epsilon$ referring to the strength of perturbation everywhere on the future event horizon. However, since $\theta_l$ and $\sigma_l$ vanishes at the bifurcation surface of the stationary black hole, they both must be of at least $\mathcal{O}(\epsilon)$ only at $\mathcal{B}$. This result is a property of the bifurcation surface itself, independent of the physical theory one considers. Hence it also generalizes beyond general relativity and holds for higher curvature theories as well. In summary we have,
$\theta_k,\, \sigma_k,\, R_{ab}k^{a}k^{b} \backsim \mathcal{O}(\epsilon)$ and $\theta_l,\sigma_l \backsim \mathcal{O}(\epsilon)$ at $\mathcal{B}$. As a result, terms like $\theta_k \theta_l \backsim \mathcal{O}(\epsilon^2)$ only at the bifurcation surface.

Having defined the geometry of the horizon, we are now set to illustrate the physical process version of first law for an arbitrary diffeomorphism invariant theory of gravity in its most general form. In order to discuss the PPFL, we need to define some suitable notion of entropy of the horizon. Since we consider theories beyond general relativity, the Bekenstein area law no longer holds. Nevertheless, whatever the expression for entropy might be, it must be some local functional integrated over the horizon. Hence we start by considering the following expression for horizon entropy:
\begin{equation}
S = \frac{1}{4} \int_{\mathcal{H}} (1+\rho) \sqrt{h}\;d^{D-2}x ,\label{BH entropy},
\end{equation}
where $\rho$ is some entropy density constructed locally on the horizon and may contain the higher curvature contributions. The area-entropy relation in general relativity limit can be obtained by setting $\rho=0$. If the black hole is stationary, the entropy should coincide with the Wald formula, but for non stationary case, can also be different. \\
The field equation in such a general theory can always be written as,
\begin{equation}
G_{ab}+H_{ab} = 8\pi T_{ab}\label{general equation},
\end{equation}
where, the term $H_{ab} $ represents the deviation from general relativity. Let us now compute the variation of the entropy along the horizon generator $k^{a}$, in response to some influx of matter,
\begin{align}
\delta S(\rho) &= \frac{1}{4}\int_{\mathcal{H}} d^{D-2}x \int \frac{d}{d\lambda}[(1+\rho)\sqrt{h}]\;d\lambda \nonumber \\
&=  \frac{1}{4}\int_{\mathcal{H}} \sqrt{h} \, d^{D-2}x \int d\lambda\, \Theta_{k},
\end{align}
where $\Theta_{k}= \theta_k + \rho \theta_k + \frac{d\rho}{d\lambda}$. In case of general relativity, $\Theta_k$ is simply the expansion of the horizon generators. Otherwise, this can only be interpreted as the change in entropy per unit area; we will call this the \textit{generalized expansion}. The entropy change can be further simplified by integrating by parts and finally the change in entropy between two cross sections at $\lambda_1 $ and $\lambda_2$ takes the form,
\begin{align}
\delta S(\rho) =&\frac{1}{4}\left(\int  dA\; \lambda \Theta_{k}\right)_{\lambda_1}^{\lambda_2} + 2\pi \int dA \;d\lambda\; \lambda\, T_{ab}k^a k^b + \nonumber \\ &\frac{1}{4}\int dA\; d\lambda\; \lambda\left[-\left(\frac{D-3}{D-2}\right)(1+\rho)\theta_k^2 +(1+\rho)\sigma^2 \right] \nonumber\\&-
\frac{1}{4} \int dA\;d\lambda\; \lambda \left(\frac{d^2\rho}{d\lambda^2} + 2\theta_k \frac{d\rho}{d\lambda} -\rho R_{ab}k^a k^b + H_{ab}k^a k^b\right).\label{general_PPFL}
\end{align}

To derive this equation, we have used Raychaudhuri equation as well as the field equation of the form $G_{ab} + H_{ab} = 8 \pi T_{ab}$. We would like to emphasize that Eq.~(\ref{general_PPFL}) represents the most general form of the variation of entropy along the null generator and no assumption regarding the strength of the perturbation or the range of integration has been made throughout the derivation. We can now discuss the change in entropy at various orders of perturbation. Since terms like $\theta_k^2,\,\sigma_k^2$ and $\theta_k (d\rho/d\lambda)$ are of $\mathcal{O}(\epsilon^2)$, they do not contribute to the first order variation. Hence, truncating the general result upto first order in perturbation, we find that the first order change in entropy takes the form,
\begin{align}
\delta S^{(1)}(\rho) =& \frac{1}{4}\left(\int dA\; \lambda \Theta_{k}\right)_{\lambda_1}^{\lambda_2} + 2\pi \int dA \;d\lambda\; \lambda T_{ab}k^a k^b \nonumber \\  -&
\frac{1}{4} \int dA\;d\lambda\; \lambda \left(\frac{d^2\rho}{d\lambda^2}  -\rho R_{ab}k^a k^b + H_{ab}k^a k^b\right).\label{firstorder}
\end{align}
Let us evaluate the last integrand of the above equation for some simplified models, say general relativity with $f(R)$ correction, for which, the full field equation takes the form,
\begin{equation}
G_{ab} + \alpha\left(f'(R)R_{\mu\nu}-\frac{1}{2}g_{\mu\nu}f(R)+g_{\mu\nu}\square f'(R) -\triangledown_\mu \triangledown_\nu f'(R) \right)= 8\pi T_{\mu\nu}.
\end{equation}
We now need an expression of the horizon entropy for $f(R)$ gravity. Let us first use the Wald entropy formula for stationary black holes Eq.~(\ref{waldf}) which gives $f'(R)$ represents the modifications to the entropy density over and above the \EH expression, i.e., $\rho= \alpha f'(R)$, where the prime denotes first derivative w.r.t the Ricci scalar $R$. Also, one can always rewrite the field equation for $f(R)$ theory in the form of Eq.~(\ref{general equation}), with 
\begin{align}
H_{ab}k^a k^b= \alpha \left(f'(R)R_{ab}k^a k^b - k^a k^b \nabla_a\nabla_b f'(R)\right).
\end{align}
Substitution of the above expression for $H_{ab}k^a k^b$ results into the following identity for $f(R)$ theories with $\rho= \alpha f'(R)$,
\begin{equation}
\int dA\;d\lambda\; \lambda \left(\frac{d^2\rho}{d\lambda^2}  -\rho R_{ab}k^a k^b + H_{ab}k^a k^b\right)=0. \label{entropy flow}
\end{equation}
Eq.~(\ref{entropy flow}) is a property of the entropy density.  If this equation is valid, our expression for the change in entropy upto first order as given in Eq.~(\ref{firstorder}) will take a very simplified form. Motivated by this result, we argue that, Eq.~(\ref{entropy flow}) could be a general property of the entropy density and holds in an arbitrary diffeomorphism invariant theory of gravity, at least up to the first order in perturbation. In fact, for $f(R)$ gravity including general relativity, it holds as an exact identity if we choose the entropy density as $\rho= \alpha f'(R)$. Hence in general we demand that Eq.~(\ref{entropy flow}) is of the form,
\begin{equation}
\int dA\;d\lambda\; \lambda \left(\frac{d^2\rho}{d\lambda^2}  -\rho R_{ab}k^a k^b + H_{ab}k^a k^b\right)= O(\epsilon^2) \label{intcond}
\end{equation}
If this is valid for a theory of gravity, the first order variation of the entropy simplifies to,
\begin{equation}
\delta S^{(1)}(\rho) =\frac{1}{4}\left(\int dA\; \lambda ~\Theta_{k}\right)_{\lambda_1}^{\lambda_2} + 2\pi \int dA \;d\lambda\; \lambda\, T_{ab}k^a k^b \label{firstorder-entropy}
\end{equation}

This is the linearised change of the entropy between two arbitrary cross sections at $\lambda = \lambda_1$ to $\lambda = \lambda_2$ provided the condition in Eq. (\ref{intcond}) holds. The first term in the $r.h.s$ is a boundary term and can be interpreted as a change of the energy $\delta E$ associated with the horizon membrane. Then, we have a version of the physical process law of the form: $T \delta S = \delta E + \delta Q$ \cite{Chakraborty:2017kob}.
Now, let us spells out two more assumptions which we like to use,

\begin{itemize}
\item The horizon possesses a regular bifurcation surface in the asymptotic past, which is set at $\lambda = 0$ in our coordinate system.

\item The horizon is stable under perturbation and eventually settle down to a new stationary black hole. So, all Lie derivatives with respect to horizon generators vanish in the asymptotic future.
\end{itemize}
The second assumption is motivated by the cosmic censorship conjecture which asserts that the black hole horizon must be stable under perturbation so that expansion and shear vanish in the asymptotic future at $\lambda = \lambda_f$. This is in principle similar to the assertion that a thermodynamic system with dissipation ultimately reaches an equilibrium state. This is a desirable property of the black hole horizon. Moreover, while deriving the physical process first law, we are already neglecting higher order terms, and this requires that small perturbations remain small throughout the region of interest.  This is equivalent to not having any caustic formation on any portion of the dynamical horizon. Under these assumptions, the boundary term doesn't contribute when integrated from the bifurcation surface to a stationary slice. This is because we have set the bifurcation surface to be at $\lambda=0$ and the final stationary cross-section, all temporal derivatives vanish. \footnote{This requires the generalized expansion $\Theta_k$ goes to zero in the future faster than the time scale $1/\lambda$ \cite{Jacobson:2003wv}. } Ultimately, what we left with is,
\begin{align}\label{BH_Dyn_03}
\delta S&= 2 \pi \int _{\lambda = 0}^{\lambda _f}\lambda ~d\lambda\, dA \,T_{ab}k^{a}k^{b}~.
\end{align}
Subsequently identifying the background Killing field as $\xi^a = \lambda\, \kappa \,k^a$ one can rewrite the above equation as,
\begin{align}
\frac{\kappa}{2\pi}\delta S=\int_{{\cal H}}  T_{ab}\,\xi^{a}\, d\Sigma^b~.
\end{align}
This completes the standard derivation of what is known as the \textit{integrated version of the physical process first law}. If the matter field satisfies the null energy condition, then one will have $T_{ab}k^{a}k^{b}\geq 0$. As a consequence, it will follow that the total change in entropy between the boundary slices is also positive semi-definite. \\

The validity of Eq.~(\ref{intcond}) is an important requirement for the validity of the physical process law. As of now, there is no general proof of the condition Eq.~(\ref{intcond}). In case $f(R)$ gravity, the condition Eq.~(\ref{intcond}) holds as an exact identity leading to the physical process law for such a theory \cite{Jacobson:1995uq}. Same can be established for Einstein Gauss-Bonnet and Lovelock class of theories \cite{Chatterjee:2011wj, Kolekar:2012tq, Sarkar:2013swa}. But, there is still no general proof of this condition. We will discuss more on this in the later sections.\\

In comparison with the equilibrium state version of the first law, the PPFL is local and independent of the asymptotic structure of the space-time. The relationship between these two versions is not straightforward. In the next section, we would like to understand how these two approaches are related to each other.
\section{Equilibrium state version and Physical process Law}

The equilibrium state version compares two nearby stationary solutions in the phase space which differ infinitesimally in ADM mass and relates the change of ADM mass $\Delta M$ to the entropy variation $\Delta S$ as,
\begin{equation}
\frac{\kappa}{2 \pi} \Delta S = \Delta M.
\end{equation}
The variation $\Delta$ is to be understood in the space of solutions. 

To understand the relationship with PPFL, consider a time-dependent black hole solution: say for simplicity, a spherically symmetric Vaidya black hole which is accreting radiation. The metric for such a space-time is \cite{Vaidya:1951zza},
\begin{equation}
ds^2 = - \left(1 - \frac{ 2 M(v) }{ r} \right)dt^2 + 2 dv \,dr + r^2 d\Omega^2.
\end{equation}
The Vaidya space-time is an excellent scenario to study the physical process first law. The area of the event horizon is increasing due to the flux of the infalling matter. The rate of change of the time-dependent mass $M(v)$ represents the energy entering into the horizon. But, although $M(v)$ is changing with time, the ADM mass of the spacetime is constant, evaluated at the spacelike infinity: $M_{\textrm{ADM}} = M(v \to \infty)$. In fact, by definition, there is no physical process which can change the ADM mass of the space time. Therefore, the relationship between the PPFL and the equilibrium state version is somewhat subtle. 

To understand the relationship, we consider the Vaidya spacetime, as a perturbation over a stationary black hole of ADM mass $m$. Therefore, we assume $M(v) = m + \epsilon \, f(v)$. The parameter $\epsilon$ signifies the smallness of the perturbation. Note that, the background spacetime with ADM mass $m$ is used only as a reference; it does not have any physical meaning beyond this. In the absence of the perturbation, the final ADM mass would be the same as $m$. Therefore, we may consider the process as a transition from a black hole of ADM mass $m$ to another with ADM mass $M_{\textrm{ADM}}$, and this allows us to relate the PPFL to the equilibrium state version.

In the case of an ordinary thermodynamic system, the entropy is a state function, and its change is independent of the path. Therefore, we can calculate the change of entropy due to some non equilibrium irreversible process between two equilibrium states by using a completely different reversible path in phase space. In black hole mechanics, the equilibrium state version can be thought as the change of entropy along a reversible path in the space of solutions, whereas the PPFL is a direct irreversible process. The equality of the entropy change for both these processes shows that the black hole entropy is indeed behaving like that of a true thermodynamic entropy \cite{Wald:1995yp}. 

Having understood the relationship between these two versions of the first law for black holes, we will now study the ambiguities of Wald's construction and how PPFL is affected by such ambiguities.\\
\section{Physical Process first law and ambiguities in black hole entropy}

The entropy of a stationary black hole with a regular bifurcation surface in an arbitrary diffeomorphism invariant theory of gravity  
is given by Wald's formula \cite{Wald:1993nt} as,
\begin{equation}
S_{W} = -2\pi\int_{\mathcal{B}} \frac{\partial L}{\partial R_{abcd}}\epsilon_{ab}\epsilon_{cd} \,\sqrt{h} \,d^{D-2}x =\frac{1}{4} \int_{\mathcal{B}} (1+\rho_w) \sqrt{h} \,d^{D-2}x\label{Wald entropy}
\end{equation}
Where $\epsilon_{ab} = k_a l_b -k_b l_a$ is the bi-normal of the bifurcation surface and $\rho_w$ represents the contribution from higher curvature terms. 

As discussed in\cite{Iyer:1994ys, Jacobson:1993vj}, the ambiguities in the Noether charge construction doesn't affect the Wald entropy in case of a stationary black hole. However, if the horizon is involved in a dynamical process, i.e., for nonstationary black holes, the Wald entropy formula no longer holds and turns out to be ambiguous up to the addition of terms of the form,
\begin{equation}
\Delta S_w = \int \Omega~dA\; ,
\end{equation}
where $\Omega =(p \theta_k \theta_l + q \sigma_k\sigma_l)$ and $\sigma_k \sigma_l = \sigma_{ab}^{k}\sigma^{ab}_l$. 
Note that, terms in $\Omega$ contains an equal number of $k$ and $l$ indices and hence combine to produce a boost invariant object, although they individually transform non-trivially under boost. The coefficients $p$ and $q$  are entirely arbitrary and can not be determined from the equilibrium state version of the first law.

Comparing Eq.~(\ref{BH entropy}) and Eq.~(\ref{Wald entropy}) and taking into account the ambiguities, we can identify $\rho = \rho_w + \Omega$. This identification essentially means that the black hole entropy for a non-stationary horizon slice can always be expressed as the expression obtained from Wald's formula plus ambiguities. So, let us define $\rho$ as the black hole entropy for higher curvature correction and $\rho_w$ as the Wald entropy. Note that the black hole entropy $\rho$ coincide with Wald entropy only in the stationary limit. Now, we would like to ask a definite question: how does the physical process law get affected by the ambiguities in the Noether charge construction? We will show that as in the case of the stationary version, the physical process law for linear perturbations is also independent of these ambiguities, provided we consider the entropy change from the past bifurcation surface to the final stationary cross-section. To see this, we write the difference between the change in black hole entropy and the change in Wald entropy up to first order in expansion and shear. A straightforward calculation using Eq.~(\ref{firstorder}) for $\rho$ and $\rho_w$ shows,

\begin{equation}
\Delta S^{(1)}(\rho) -\Delta S^{(1)}(\rho_w) = \frac{1}{4}\int dA \lambda \left( \frac{d\Omega}{d\lambda} +\Omega\theta_k \; \right)\Bigg|_{\lambda_1}^{\lambda_2}- \frac{1}{4} \int dA\, d\lambda \left(\lambda \frac{d^2\Omega}{d\lambda^2}\right),
\end{equation}
where we have neglected the terms $\Omega \, R_{ab} k^a k^b$ and $\Omega \,\theta_k$ which are of $\mathcal{O}(\epsilon^2)$ and do not contribute to the first order variation. Simplifying it further one can obtain,
\begin{equation}
\Delta S^{(1)}(\rho) -\Delta S^{(1)}(\rho_w) = \frac{1}{4}\int dA\; \Omega \Big|_{\lambda = 0}^{\lambda_f}.\label{entropy difference}
\end{equation}
The above equation represents the difference in the change in black hole entropy and Wald entropy as a boundary term evaluated between  the bifurcation surface at $\lambda = 0$ and the final stationary cross section at $\lambda = \lambda_f$. Then, as discussed in the previous section, terms like $\theta_k \theta_l$ are of second order in perturbation and therefore $\Omega$ turns out to be $\mathcal{O}(\epsilon^2)$ and does not contribute to the linear order calculation. The contribution from the upper limit also vanishes as the expansion $\theta_k$ is zero on a future stationary cross-section. Hence, up to first order, the ambiguities does not affect the PPFL when integrated from a bifurcation surface to a stationary slice. This is analogous to the case of the equilibrium state version of the first law as proven in \cite{Iyer:1994ys, Jacobson:1993vj}. The integrated version of the physical process law and therefore the net change of the entropy is independent of the ambiguities in the Wald entropy construction.\\

In summary, given a particular theory, if there is a choice of entropy density $\rho$ which obeys the condition Eq.~(\ref{intcond}), then $\rho + \Omega$ will also obey the same condition. So, Eq.~(\ref{intcond}) is independent of the ambiguities as long as we the integrating from a past bifurcation surface to a stationary future cross-section. If it holds for $\rho_w$, it will hold for $\rho$ also.

This result, however, doesn't hold when second-order perturbations are considered. Unlike first order, the difference in the change in black hole entropy and Wald entropy is given by a boundary term and a bulk integral. As a result, any conclusion about the change of black hole entropy beyond linearized perturbation requires the resolution of these ambiguities. \\
Similarly, if we demand an instantaneous second law, such that the entropy is increasing at every cross-section, to hold beyond general relativity, we need to fix the ambiguities and find the appropriate black hole entropy \cite{Bhattacharjee:2015yaa, Bhattacharjee:2015qaa}. Then, it is also possible to study the higher order perturbations and obtain the transport coefficients related to the horizon \cite{Fairoos:2018pee}. \\

\section{Linerized version of the second law}

We have seen in the last section that the integrated version of the physical process law is insensitive to the ambiguities in the Wald entropy. Therefore, to fix the ambiguities, let us consider the linearized version of the second law, where we seek to evaluate the instantaneous change of the entropy due to the flux of matter. To start, we consider the expression of the change of the entropy,
\begin{align}
\Delta S(\rho) = \frac{1}{4}\int_{\mathcal{H}} \sqrt{h} \, d^{D-2}x \int d\lambda\, \Theta_{k}.
\end{align}

If we can prove that the generalized expansion is always positive on any cross-section of the horizon, we have an instantaneous increase theorem for the black hole entropy. Before proceeding with the calculation, we ponder over the implication of such a result. In ordinary thermodynamics, the entropy is generally defined for an equilibrium state. So, it makes sense to obtain the change of entropy between two equilibrium states of the thermodynamic system. The integrated version of the physical process law is an equivalent calculation for black holes. But, the area theorem in GR shows that the area/entropy is a locally increasing function, and there is a local version of the second law at a non stationary cross-section of the black hole horizon. This is indeed stronger than the global increase of the entropy. Using the local second law, we can define an entropy current associated with the horizon which has positive divergence. The existence of such a current may imply a hydrodynamical picture of the black hole mechanics as envisaged in fluid gravity duality \cite{Bhattacharyya:2008xc}.\footnote{ We thank Shiraz Minwalla to suggest this.}\\

To present the derivation, first consider the general relativity, for which $\Theta_k = \theta_k / 4$. Then, the Raychaudhuri equation and Einstein equation imply,
\begin{align}
\frac{d \Theta_{k}}{d\lambda} = - 2 \pi \, T_{ab} k^a k^b + \mathcal{O}(\epsilon^2),
\end{align}
where the higher order terms involve the squares of expansion and shear of the horizon generators. The matter flux itself is of $\mathcal{O}(\epsilon)$. So, if the matter obeys null energy condition, i.e. $T_{ab} k^a k^b > 0$ and the higher order terms are essentially small, then we have, $ d \Theta_{k} / d\lambda < 0$. So, the expansion is decreasing at every cross section. Next, we recall the assumption about the stability of the black hole, which asserts that $\Theta_k \to 0$ in the asymptotic future. This boundary condition immediately gives $ \Theta_k \geq 0$ at every slice on the horizon, and the equality holds only in the asymptotic stationary future. As a result, the area is increasing locally on the horizon. Note that importance of the boundary condition to derive the result. The assumption that the expansion vanishes in the asymptotic future ensures the stability of the black hole under perturbation. There are several aspects to this assumption. This can be argued using the Penrose's theorem that the generators of the event horizon have no end future point and as a result, there is no caustic in the future. If the expansion is negative at any instant, it will further decrease and ultimately will lead to a caustic invalidating the Penrose's result  \cite{Townsend:1997ku}. This is also related to the cosmic censorship hypothesis \cite{Wald:1984rg}.\\

We want to generalize the same result to a theory of gravity with higher curvature correction terms. To do this, let us write the black hole entropy for a non stationary cross section as $ \rho = \rho_w +  p\, \theta_k \theta_l + q\, \sigma_k\sigma_l $. Then, the evolution equation of the generalized expansion at the linearized order of the perturbation becomes

\begin{align}
\frac{d \Theta_{k}}{d\lambda} = - 2 \pi \, T_{ab} k^a k^b +  \frac{1}{4}\left(\frac{d^2\rho}{d\lambda^2}  -\rho R_{ab}k^a k^b + H_{ab}k^a k^b\right) + \mathcal{O}(\epsilon^2).
\end{align}

While discussing the integrated version, we have shown that the integrated version of physical process first law is independent of the ambiguities of Wald entropy. This is because the integral of the second term in r.h.s of the above equation is of higher order when we integrate between past bifurcation surface to the future stationary slice. So, if $\rho$ obeys the integrated version of the first law so does $\rho_w$ as both of these obey Eq.~(\ref{intcond}). On the other hand, the formulation of the local first law requires the integrand of Eq.~(\ref{intcond}) to be of higher order. Therefore, the condition for the validity of the linearized increase law is,
\begin{align}
\frac{d^2\rho}{d\lambda^2}  -\rho R_{ab}k^a k^b + H_{ab}k^a k^b = \mathcal{O}(\epsilon^2).\label{condition_local}
\end{align}
Therefore, we set the first order part of the l.h.s of the above equation to zero and determine the ambiguity coefficients $p$ and $q$. This is just one condition, but at the linearized order, expansion and shear are independent of each other, and as a result, we have two independent conditions from the above equation. As an example, consider a theory of gravity of gravity described the Lagrangian,
\begin{align}
L = \frac{1}{16 \pi} \left( R + \beta\, R_{ab} R^{ab}\right).
\end{align}

The Wald formula in Eq.~(\ref{Wald entropy}) for this theory gives $\rho_w = - 2 \beta R_{ab} k^a l^b$. Then the requirement of the validity of the condition Eq.~(\ref{condition_local}) will give  $ p = - 1/2$ and $ q = 0$ \cite{Bhattacharjee:2015yaa} and the black hole entropy becomes,

\begin{align}
S = \frac{1}{4}\int dA \left[ 1 - 2\beta \, \left(R_{ab}k^a l^b - \frac{1}{2} \theta_k \theta_l \right)\right].
\end{align}
The evolution equation of the generalized expansion for this entropy is,
\begin{align}
\frac{d \Theta_{k}}{d\lambda} = - 2 \pi \, T_{ab} k^a k^b +  \mathcal{O}(\epsilon^2).
\end{align}
This equation is exactly analogous to the linerized Raychaudhuri equation in general relativity. This is a nontrivial result; the entropy function has modified due to higher curvature corrections, the field equation is also different. But the evolution equation of the entropy for linerized perturbation remains same in form! Therefore, if we use the future stability condition i.e. $\Theta_k \to 0$ in the future, we have the instantaneous increase of the entropy at every cross section of the horizon, $\Theta_k \left(\lambda \right) > 0$ provided the matter obeys the null energy condition.\\

Similarly, consider a more general theory of gravity in $D$ dimensions with the Lagrangian,

\begin{align}
L = \frac{1}{16 \pi} \left( R + \alpha\,R^2 + \beta  \, R_{ab} R^{ab} + \gamma \, {\cal L_{GB}} \right),
\end{align}
where ${\cal L_{GB}} = R^2 - 4  R_{ab} R^{ab} +  R_{abcd} R^{abcd}$, the so called Gauss Bonnet correction term. The black hole entropy for such a theory can be obtained by fixing the ambiguities using the linerized second law and the result is,

\begin{align}
S = \frac{1}{4}\int dA \left[ 1 +  \left( 2\alpha R- 2 \beta\, \left(R_{ab}k^a l^b - \frac{1}{2} \theta_k \theta_l \right)+2 \gamma\, ^{(D-2)}R \right)\right], \label{ent_GL}
\end{align}
where $^{(D-2)}R$ is the intrinsic Ricci scalar associated with the horizon cross section.\\
This can be generalized to any theory, and it is always possible to fix the ambiguities from the linearized second law so that we have a local increase theorem at every cross-section of the nonstationary event horizon \cite{Bhattacharjee:2015yaa, Bhattacharjee:2015qaa, Wall:2015raa}. \\

If we now consider that the black hole is in an asymptotically Anti De Sitter space-time, after fixing the ambiguities, the black hole entropy becomes identical in form to the holographic entanglement entropy of the boundary gauge theory \cite{Bhattacharjee:2015yaa, Bhattacharjee:2015qaa}. The holographic entanglement entropy \cite{Ryu:2006bv} is a proposal which relates the entropy of the boundary gauge theory with the area of certain minimal surfaces in the bulk (which obeys Einstein's equation) within the context of gauge-gravity duality. The original principle has been generalized to higher curvature theories \cite{Casini:2011kv, Bhattacharyya:2014yga, Dong:2013qoa} and the entanglement entropy density of the boundary theory is given as, $\rho = \rho_w + a\, \theta_k \theta_l + b\, \sigma_k\sigma_l $,. The part $\rho_w$ is of the same form of Wald entropy for black holes, and the coefficients $a$ and $b$ depends on the choice of gravity theory in the bulk; for general relativity $ a = b = 0$. The expansions and shears correspond to that of a codimension two surface which is anchored to a region of the boundary.  Note, a priory, this entanglement entropy is not related to the entropy of the black hole in the bulk. Also, this entropy has no ambiguities, and the coefficients $a$ and $b$ can be calculated using AdS-CFT \cite{Dong:2013qoa}. Our calculations show, if we consider a nonstationary black hole in the bulk and demand that the black hole entropy obeys linearized second law, we will have $ p = a$ and $ q = b$ \cite{Bhattacharjee:2015qaa}. It is indeed remarkable that the entropy for black holes in AdS spacetime which obeys linearized second law turns out to be related with the holographic entanglement entropy. It seems that the validity of black hole thermodynamics is already encoded in the holographic principle; the holographic entanglement entropy satisfies the linearized second law while the Wald entropy does not. \\

Let us summarise the main results: a theory of gravity which has black hole solutions will obey the integrated version of the physical process law if Eq.~(\ref{intcond}) holds. Given a theory and an expression of black hole entropy, we can always verify the validity of this condition. Also, the condition Eq.~(\ref{intcond}) is independent of the ambiguities of the Noether charge construction, as long as we are integrating from initial bifurcation surface to a future stationary cross-section. Therefore, in any theory, if Wald entropy $\rho_w$ satisfied this condition, so does the black entropy $\rho$. On the other hand, the local increase law depends on the validity of  Eq.(~\ref{condition_local}) which is sensitive to the ambiguities. Hence, there is only a particular choice of the ambiguity coefficients $p$ and $q$ for which the local increase law for linearized fluctuations holds. Remarkably, such a choice makes the black hole entropy identical in form to holographic entanglement entropy.

\section{Beyond the linearized second law}

The next obvious question is to find the full evolution equation of the entropy. For general relativity, the full Raychaudhuri equation Eq.~(\ref{rc}) with null energy condition still gives $d\theta_k / d\lambda <0$ and this leads to the area theorem. Beyond general relativity, the calculation is straightforward and gives an evolution equation for the generalized expansion $\Theta_k$. We will only present the final result, for more details about the derivation, refer to \cite{Fairoos:2018pee}. We consider a $D$ dimensional Einstein-Gauss-Bonnet theory; the entropy is then given by,

\begin{align}
S = \frac{1}{4}\int dA \left( 1 +  2 \gamma\, ^{(D-2)}R \right).
\end{align}

This entropy can be obtained by setting $ \alpha = \beta = 0$ in the expression Eq.~(\ref{ent_GL}). Then, the full evolution equation of the generalized expansion is,

\begin{equation}
\begin{gathered}
4  \frac{d\Theta_k}{d\lambda}=-\frac{\theta^{(k)2}}{D-2}-\sigma^{(k)ab}\sigma^{(k)}_{ab}-6\gamma\frac{(D-4)\theta^{(k)2}\mathcal R}{(D-2)^2}-2\gamma\sigma^{(k)ab}\sigma^{(k)}_{ab}\mathcal R\\-4\gamma\frac{(D-8)\theta^{(k)}\sigma^{(k)ab}\mathcal R_{ab}}{(D-2)}
+8\gamma\sigma^{(k)a}_c\sigma^{(k)cb}\mathcal R_{ab}-4\alpha \mathcal R_{fabp}~\sigma^{(k)ab}\sigma^{(k)pf}\\
+2\gamma\Bigg[2\bigg(D_c\beta^c\bigg)\bigg(K^{(k)}_{ab}K^{(k)ab}\bigg)-4\bigg(D_c\beta^b\bigg)\bigg(K^{(k)}_{ab}K^{(k)ac}\bigg)+2\beta^c\beta_cK^{(k)}_{ab}K^{(k)ab}\\-4\beta_cK^{(k)}_{ab}\beta^bK^{(k)ac}\Bigg]
+4\gamma \bigg[2\bigg(D^b\beta^f\bigg)\bigg(K^{(k)}K^{(k)}_{bf}\bigg)-2\bigg(D_a\beta^a\bigg)\bigg(K^{(k)}\bigg)^2\\+ 2h^{ab}\beta^cK^{(k)}_{ac}\beta_bK^{(k)}-h^{ab}\beta_a\beta_b(K^{(k)})^2\bigg]
+4\gamma R_{kk}\frac{(D-3)(D-4)\theta^{(n)}\theta^{(k)}}{(D-2)^2} \\-4\gamma h^{ac}h^{bd}R_{kckd}\frac{(D-4)\theta^{(k)}\sigma^{(n)}_{ab}}{D-2}-4\gamma h^{ac}h^{bd}R_{kckd}\frac{(D-4)\theta^{(n)}\sigma^{(k)}_{ab}}{D-2}\\
+8\gamma h^{ac}h^{bd}R_{kckd}\sigma^{(k)}_{af}\sigma^{(n)f}_b-4\gamma R_{kk}\sigma^{(k)}_{ab}\sigma^{(n)ab}-8\pi G\, T_{kk}
\\+\gamma\, \text{(total derivatives)}. \label{fullexp}
\end{gathered}
\end{equation}
\begin{equation*}
\begin{gathered}
\end{gathered}
\end{equation*}

To comprehend this formidable equation, let us first spell out the notations. $K^{(i)}_{ab}$ is the extrinsic curvature of the horizon cross section w.r.t the null normal $i = k , l$. We have also used the notations $R_{k c  k d} = R_{\mu c  \rho d } k^\mu k^\rho$, $ \beta_a = - l^\mu \grad_a k_\mu$ etc. Setting $ \gamma= 0$, we will obtain the familiar null Raychaudhuri equation. Otherwise, this equation is the thermodynamics generalization of the null Raychaudhuri equation. The expansion and shear of the horizon generators, i.e., $\theta^{(k)}$ and $\sigma^{(k)}_{ab}$ vanish on the background stationary horizon and therefore are at least linear order in perturbation. But, the expansion of the auxiliary null vector $\theta^{(l)}$ is non zero even on the stationary horizon. The total derivative terms involve spacial derivative of the extrinsic curvatures and are second order in perturbation. If we consider only the terms linear in perturbation, we will obtain:
\begin{equation}
\frac{ d \Theta_k}{d \lambda} = - 2 \pi \, T_{ab}k^a k^b + {\cal O}(\epsilon^2).
\end{equation}
This is the equation which will give us the linearized version of the physical process law. \\

The full equation will give an exact expression of the change of horizon entropy. We would like to apply this equation to understand the full evolution of the horizon entropy. Due to the complicated structure of the terms, it is difficult to obtain any conclusion in general. So, to make sense of this equation, we will now specialize to the case of spherically symmetric second-order perturbations about a static black hole background with maximally symmetric horizon cross-section \cite{Bhattacharjee:2015yaa, Bhattacharjee:2015qaa}. Then, in a \textit {order by order} calculation in  $\theta^{(k)}$ and $\sigma^{(k)}_{ab}$, we will obtain followings up to second order: 

\begin{equation}
\frac{ d \Theta_k}{d \lambda} = - 2 \pi \, T_{ab}k^a k^b - \zeta \, \theta_{k}^{2},
\end{equation}
where the quantity $\zeta$ is to be evaluated on the background horizon. There is no shear because we have assumed spherically symmetric perturbation only. Now consider a situation where the stationary black hole is perturbed by some matter flux, and we are examining the second law when the matter has already entered into the black hole. In that case, the above evolution equation does not have any contribution from matter stress-energy tensor and the evolution will be driven solely by the $\theta_{k}^{2}$ term.  In such a situation, if we demand the entropy is increasing, we have to fix the sign of the coefficient of  $\theta_{k}^{2}$; the quantity $\zeta$. We evaluate the coefficient in the stationary background and impose the condition that overall sign in front of  $\theta_{k}^{2}$ is negative. This will immediately give us a bound on the parameters of the theory under consideration.
To illustrate this, we now consider specific cases. First consider the case when the background is a spherically symmetric solution of the Einstein Gauss-Bonnet (EGB) gravity with metric,
\begin{equation}
ds^2 = -f(r) \, dt^2 + \frac{dr^2}{f(r)} + r^2 \, d\Omega^{2}_{D-2}.
\end{equation}
The expression of $\zeta$ is now given by,
\beq 
\zeta=\frac{1}{D-2}+\frac{(D-4)\gamma}{(D-2)^2} \left[6\,^{(D-2)}R - \frac{2\,(D-3)(D-2) f'(r)}{ r(v)} \right]. \label{zeta}
\eeq 
As discussed earlier, we will evaluate $\zeta$ for different backgrounds and determine bounds on the coefficient $\gamma$ from the constraint $\zeta > 0$. For the EGB gravity, we first consider the $5$-dimensional spherically symmetric, asymptotically flat Boulware-Deser (BD) \cite{Boulware:1985wk}  black hole as the background, for which the horizon radius $r_h$ is related to the mass $M$ as, $r_{h}^{2}+2\gamma=M$ and the existence of an event horizon demands $r_h^2 > 0$. Now, evaluating $\zeta$ for the above background at the horizon $r =r_{h}$, and imposing that $\zeta > 0$, we obtain the condition, $M>2 |\g|\, \textrm{if}~ M>0$. To understand this better, note that we require $M>2\g$ to avoid the naked singularity of the black hole solution for $\g>0$. Thus in this case for a spherically symmetric black hole, $\zeta$ will be positive and hence second law will be automatically satisfied. The condition of the validity of the second law is same as that for having a regular event horizon.
Also, for $\g > 0$, it is possible to make $r_h$ as small as possible by tuning the mass $M$. But when $\g$ is negative (a situation that appears to be disfavoured by string theory, see \cite{Boulware:1985wk}, \cite{bms} and references therein), $r_h$ cannot be made arbitrarily small, and it would suggest that these black holes cannot be formed continuously from a zero temperature set up. Notice that we could have concluded the same without the second law if $M$ is considered to be positive--however, our current argument does not need to make this assumption. Due to this pathology, it would appear that the negative Gauss-Bonnet coupling case would be ruled out in a theory with no cosmological constant. \\

The case for the $5$-dimensional AdS black hole solution for EGB gravity with cosmological constant $ \Lambda = - (D-1)(D-2)/2 l^2  $ as the background is more interesting. Now the horizon could be of planar, spherical or hyperbolic cross sections. We will first consider a black brane solution with a planar horizon. Then we obtain $\zeta= 1/(D-2)\left(1-2(D-1)\lambda_{GB}\right)$ where we have introduced a rescaled coupling in $D$ dimensions as $\lambda_{GB} l^2=(D-3)(D-4)\g$. Again demanding positivity of $\zeta$ we get \cite{Bhattacharjee:2015qaa},
\beq \label{planarbound}
\lambda_{GB}< \frac{1}{2(D-1)}\,.
\eeq
Remarkably, in $D=5$ this coincides with the bound which has to be imposed to avoid instability in the sound channel analysis of quasi-normal modes of a black hole in EGB theory which is taken to be the holographic dual of a conformal gauge theory. It was shown in \cite{Buchel} that when $\lambda_{GB} > 1/8$ the Schroedinger potential develops a well which can support unstable quasi-normal modes in the sound channel.  It is quite interesting to see that the second law knows about this instability.  \\
Another interesting case corresponds to the hyperbolic horizon. In this case, the intrinsic scalar is negative, and if we also assume that $\g > 0$, then there is an obvious bound on the higher curvature coupling beyond which the entropy itself becomes negative and thereby loses any thermodynamic interpretation. This bound in general $D$ dimension is found as $\lambda_{GB} < D(D-4) / 4 (D-2)^2$. If the analysis of the second law has any usefulness, it must provide a more stringent bound for the coupling $\g$, and that is indeed the case. Also, to analyze the case for hyperbolic horizons, we will only consider the so-called zero mass limit. In the context of holographic entanglement entropy, these topological black holes play an important role as shown in \cite{Myers:2010xs, Myers:2010tj, Casini:2011kv}. One can relate the entanglement entropy across a sphere to the thermal entropy in $R \times H^{D-2}$ geometry by a conformal transformation. \\
Now for holographic CFTs, one has to evaluate the Wald entropy for these topological black holes as they are dual to the field theory placed on $R \times H^{D-2}$ to obtain the entanglement entropy across a spherical region at the boundary.  In our context, imposing $\zeta > 0$, it turns out that the zero mass limits gives the most stringent bound on the coupling $\lambda_{GB}$ given by \cite{Bhattacharjee:2015qaa},
\beq
\lambda_{GB}< \frac{9}{100}. \label{GBbound}
\eeq
First, note that this bound on $\lambda_{GB}$ is independent of the dimensions. Also, comparing with the bound in Eq.~(\ref{planarbound}) we can easily see that up to $D=6$, the bound in Eq.~(\ref{GBbound}) is strongest but from $D=7$ onwards Eq.~(\ref{planarbound}) is the strongest one. 
Next, in the five dimensions, the bound in Eq.~(\ref{GBbound}) quite curiously coincide with the tensor channel causality constraint \cite{hm, Brigante, holoGB}. For $D>5$, this bound Eq.~(\ref{GBbound}) from the second law will be stronger than the causality constraints. \\
In principle, it is possible to repeat this analysis for any higher curvature gravity theory to obtain similar bounds on the higher curvature couplings provided we have an exact stationary black hole solution as the background \cite{Bhattacharjee:2015qaa}. These bounds will be necessary if we demand that the second law of thermodynamics holds for an observer outside the horizon. Any quantum theory of gravity which reproduces such higher curvature corrections and also aims to explain the microscopic origin of black hole entropy must satisfy these bounds. We can constrain various interesting gravity theories in $4$ dimensions by our method. In $4$ dimensions, our method is the only one to constrain these theories where the causality based analysis \cite{Camanho:2014apa} is insufficient. For example, for critical gravity theories in $D=4$ \cite{Lu:2011zk} analyzing black holes in AdS background we obtain the bound on the coupling ($\alpha_{c}$), $-\frac{1}{2}\leq \alpha_{c} \
\leq \frac{1}{12}.$ Also, for New Massive gravity in $D=3$ \cite{Bergshoeff:2009hq, Grumiller:2009sn} we obtain the bound on couplings ($\sigma$) as, $-3\leq \sigma\leq \frac{9}{25}$.\footnote{The lower bound for both these two cases are coming from demanding the positivity of the entropy.} 

In conclusion, these results show that the validity of a local increase law of black hole entropy can constrain the parameters of the higher curvature terms. Any theory of gravity which does not obey these bounds will have a severe problem with the second law in the presence of a black hole. \\

Interestingly, there are works which suggest that the higher curvature gravity does not make sense as a stand-alone classical theory. Consider Einstein-Gauss-Bonnet gravity in dimensions greater than four. The theory has exact shock wave solutions which can lead to a negative Shapiro time delay. This can be used to create a time machine: closed timelike curve without any violation of energy conditions \cite{Camanho:2014apa}. As a result, such higher curvature theories have badly behaved causal properties for either sign of the higher curvature coupling. Hence, it is proposed that these theories can only make sense as an effective theory and any finite truncation of the gravitational action functional will lead to pathological problems. This result is criticized in \cite{Papallo:2015rna} where gravitons propagating in smooth black hole spacetimes are considered. It is shown that for a small enough black hole, the gravitons of appropriate polarisation, and small impact parameter, can indeed experience negative time delay, but this can not be used to build a time machine. This is because the required initial data surface is not everywhere space like and therefore the initial value problem is not well-posed. Nevertheless, the result of \cite{Camanho:2014apa} is quite significant and needs careful understanding.\\

Similar conclusions can be obtained about the validity of the classical second law for black hole mergers in Lovelock class of theories \cite{Liko:2007vi, Sarkar:2010xp}. In such theories, it is possible to construct scenarios involving the merger of two black holes in which the entropy instantaneously decreases. But, it is also argued that the second law is not violated in the regime where Einstein-Gauss-Bonnet theory holds as an effective theory and black holes can be treated thermodynamically \cite{Chatterjee:2013daa}.

\section{ Conclusions and open problems}
Black hole thermodynamics provides a powerful constraint on any proposal to understand the quantum gravitational origin of black hole entropy. The area law has motivated significant progress in theoretical physics; most importantly the holographic principle.  Similarly, the pioneering work by Jacobson \cite{Jacobson:1995ab} where he considered the concept of local Rindler horizons and showed that Einstein field equations could be derived from thermodynamic considerations hints a deep thermodynamic origin of the full dynamics of gravity. \\

Similar results are proven in a more general context by Padmanabhan and collaborators. They have shown that the field equation of any higher curvature gravity theory admits an intriguing thermodynamic interpretation \cite{Padmanabhan:2009jb, Padmanabhan:2009vy}. Interestingly, the result is also valid beyond black hole horizons and for any null surface in space-time \cite{Chakraborty:2015aja}. These fascinating results lead an alternative approach ``the emergent gravity paradigm" to understand the dynamics of gravity \cite{Padmanabhan:2014jta}. There is also a local gravitational first law of thermodynamics formulated using the local stretched light cones in the neighbourhood of any event \cite{Parikh:2018anm}. This result indicates that certain geometric surfaces - stretched future light cones - which exist near every point in every spacetime, also behave as if they are endowed with thermodynamic properties. All these results seems to suggest that the thermodynamic properties of space time transcends beyond the usual black hole event horizon. \\

The derivation of a full second law beyond general relativity remains an important open problem. Ideally, we would like to follow a nonperturbative approach and find a suitable generalization of the area theorem with some restriction on the higher curvature parameters. This requires understanding the thermodynamic Raychaudhuri equation like Eq.~(\ref{fullexp}) for an arbitrary theory of gravity. This is a formidable but straightforward problem. We also like to understand the relationship between holographic entanglement entropy and black hole entropy. The area theorem may have some interesting holographic interpretations. The Holographic Entanglement Entropy was shown to obey various nontrivial inequalities. One of these is the strong subadditivity condition (SSA) which is a fundamental property of entanglement entropy in any quantum field theory and a central theorem of quantum information theory. It is known that the violation of SSA for the boundary theory is connected with the violation of the null energy condition in the bulk spacetime \cite{Allais:2011ys, Callan:2012ip, Caceres:2013dma}. Since null energy condition is a requirement for the validity of the Hawking area theorem, it is expected that there exists a strong connection between the area theorem for black holes and SSA for holographic entanglement entropy. This relationship may provide us a better understanding of the scope and applicability of the holographic principle. \\

We end this review with a quotation by Arthur Eddington, \\

\textit{``The law that entropy always increases, holds, I think, the supreme position among the laws of Nature. If someone points out to you that your pet theory of the universe is in disagreement with Maxwell's equations - then so much the worse for Maxwell's equations. If it is found to be contradicted by observation - well, these experimentalists do bungle things sometimes. But if your theory is found to be against the second law of thermodynamics I can give you no hope; there is nothing for it but to collapse in deepest humiliation."}\\

Same can be said for any theory of gravity which has a black hole solution. 

\section{Acknowledgements} 

This review is based on the work done in collaboration with Aron Wall, Srijit Bhattacharjee, Arpan Bhattacharyya, Aninda Sinha, Fairoos C, Akash K Mishra, Avirup Ghosh, Sumanta Chakraborty, and Maulik Parikh.  The author thanks Amitabh Virmani for his encouragement to write this review. Special thanks to T Padmanabhan, Ted Jacobson, Aron Wall, Aninda Sinha and Maulik Parikh for sharing their deep insights about black hole physics. SS also acknowledge many constructive comments from the referees on the previous draft of this review. The research of SS is supported by the Department of Science and Technology, Government of India under the Fast Track Scheme for Young Scientists (YSS/2015/001346).

\end{document}